\documentstyle[10pt,aaspp4]{article}

\begin{document}

\def\simg{\mathrel{%
      \rlap{\raise 0.511ex \hbox{$>$}}{\lower 0.511ex \hbox{$\sim$}}}}
\def\siml{\mathrel{%
      \rlap{\raise 0.511ex \hbox{$<$}}{\lower 0.511ex \hbox{$\sim$}}}}
\def\Mesz{M\'esz\'aros }
\def\ie{i.e$.$~} \def\eg{e.g$.$~} \def\etal{et al$.$~}

\begin{center}
\title{\sc Radiative Regimes in Gamma-Ray Bursts and Afterglows}
\author{A. Panaitescu \& P. M\'esz\'aros}
\affil{Department of Astronomy \& Astrophysics,\\
    Pennsylvania State University, University Park, PA 16802}
\end{center}

\begin{abstract}
 We present numerical simulations of Gamma-Ray Bursts arising from external 
shocks in the impulsive and wind models, including a weak or a strong 
coupling between electrons and protons plus magnetic fields, and analyze 
the burst features in each scenario. The dynamics of the ejecta and external 
medium are followed into the late stages of deceleration, in order to study
the hydrodynamics of the remnant and the temporal and spectral evolution of 
the afterglow. A brief comparison with the optical and radio afterglows
of GRB 970228 and GRB 970508 is made.
\end{abstract}


\section{Introduction}

Gamma-Ray Bursts (GRBs) are thought to be due to internal or external shocks
in relativistic fireball outflows following a catastrophic compact binary
merger or collapse. This view has received considerable support from recent
observations of GRB afterglows in X-ray, optical and radio, extending in some
cases over many months (\eg Proceedings of the Fourth Huntsville GRB 
Symposium -- Meegan, Preece \& Koshut 1998). While the $\gamma$-ray emission 
of some bursts, particularly those with very many peaks, probably arises 
from internal shocks (Rees \& \Mesz 1994, Sari \& Piran, 1997), 
these should be followed in most cases by external shocks.
The simplest afterglow model is provided by the time evolution
of the decaying external shock (\Mesz \& Rees 1997), 
and can explain the major features of observed afterglows (Wijers, Rees \& 
\Mesz 1997, Tavani 1997, Vietri 1997, Waxman 1997a, Reichart 1997).
In this paper, we simulate GRB afterglows
in the framework of the external shock model (\Mesz \& Rees 1993), whose
$\gamma$-ray light curves are either fairly smooth or have a low ($\siml 10$) 
number of pulses (Panaitescu \& \Mesz, 1998a). Our purpose is to investigate
the properties of bursts and afterglows under different physical conditions 
which impact the dynamic regime of expansion of the 
remnant as well as the burst and afterglow spectrum.

The details of the hydrodynamic code and of the energy release and transfer 
model used here (including assumptions and approximations) are presented in 
Panaitescu \& \Mesz (1998a). Here we mention only the most important 
assumptions and the new features included : \\
\hspace*{3mm} 1. The electrons are initially accelerated by one of the two shocks
    (``reverse" and ``forward") that sweep up the relativistic ejecta or 
    the external medium, respectively. The distribution of electrons is a power-law  
    of index $p=2.5$, from a minimum Lorentz factor $\gamma_m$ to a maximum 
    $\gamma_M$. The $\gamma_m$ is derived from the total energy of electrons, assumed 
    to be a fraction $\varepsilon_{el}$ of the internal energy of the shocked fluid: 
    $\gamma_m = 610\, \varepsilon_{el} \Gamma$, where $\Gamma$ is the flow Lorentz
    factor. $\gamma_M$ is  upper bounded by the condition that electrons 
    with this Lorentz factor can be accelerated on timescale shorter than their
    cooling timescale. \\
\hspace*{3mm} 2. The electrons lose energy through synchrotron emission in the 
  presence of a turbulent magnetic field, and through inverse Compton of the 
   self-produced synchrotron photons. The magnetic field intensity $B'$ is 
   parameterized by the
   fraction $\varepsilon_{mag}$ of the internal energy stored in the magnetic 
   field: $B'^2/8\pi = \varepsilon_{mag} n'_e m_p c^2 \Gamma$, where $n'_e = 
   4 \Gamma n_{ext}$ is the co-moving density of the shocked fluid, $n_{ext}$ being
   the number density of the decelerating medium.  At equipartition between 
   electrons, protons and the magnetic field, $\varepsilon_{el}= 
   \varepsilon_{mag}=1/3$. \\
\hspace*{3mm} 3. In the {\it weak coupling model} the energy transfer from protons 
  and magnetic field to electrons takes place on a hydrodynamic (deceleration) 
  timescale $t_{dec}$.  In the {\it strong coupling model} electrons are 
  assumed to be re-accelerated on a timescale much shorter than $t_{dec}$ (\eg
  by repeated scatterings on magnetic field inhomogeneities, or other mechanisms). 
  These are two extreme situations; in general such re-accelerations should occur 
  on an intermediate timescale that must be treated as a new free parameter, due 
  to the relatively poor present understanding of the microscopic processes that 
  could be at work here. \\
\hspace*{3mm} 4. Unlike in our previous paper, we use now the full shape of the 
  synchrotron spectrum to calculate the emission from the two shocks.
    However, for higher computational efficiency, we maintain the previous 
    ``monochromatic approximation" when calculating the inverse Compton (IC) 
    spectrum: before up-scattering, the spectrum of the synchrotron radiation 
   from each shock is approximated as monochromatic, at an intensity-weighted 
   frequency. Furthermore, the IC spectrum from the electrons in each infinitesimal 
    volume element is approximated as monochromatic, at the peak frequency for 
    a given electron Lorentz factor and energy of incident synchrotron photon. \\
\hspace*{3mm} 5. Self-absorption of low energy (radio and optical) photons and 
   the destruction of high energy photons ($\simg 1\,{\rm GeV}$) through pair 
   creation during propagation from source to observer are not taken into account.
    We consider cosmological effects, assuming that the source is located at 
   redshift $z=1$ and that $H_0 = 75\, {\rm km\,s^{-1} Mpc^{-1}}$ and $\Omega=1$.

The interaction between the expanding shell and the external matter is simulated 
using a 1D hydrodynamical code suitable for relativistic flows involving shocks. 
The temporal features (\ie bolometric and band light-curves) and spectral features 
(a set of instantaneous spectra and the averaged spectrum) of the burst are 
calculated by integration over lab-frame time, volume of the shocked fluid and 
electron distribution. As pointed out by Waxman (1997b), most of the radiation 
received by the observer at given time $T$ comes from a ring whose width is 
relatively small compared to the size $\sim \Gamma(T)\, cT$ of the visible disk, 
where $\Gamma(T)$ is the Lorentz factor of the fluid moving exactly toward the 
observer. We found that the shape of the source, as seen by the observer in a 
given band, changes from an almost uniformly bright disk to a ring whose width 
(defined as the on-sky projected size of the zone that radiates 50\% of the energy
received at detector) is between 6\% and 21\% of the radius of the source's 
projection on the sky (Panaitescu \& \Mesz 1998b). The fluid seen in this ring 
is shocked earlier than the fluid moving with $\Gamma(T)$ on the line of sight 
toward the center of explosion, and thus they have different physical parameters 
($\gamma_m$, $B'$, $\Gamma$). This is taken into account in the analytic 
calculations presented in \S 3.

\section{Radiative Dynamics and Gamma-Ray Emission}

We investigate here the effect of a continuous energy transfer between protons  
plus magnetic fields and the radiating electrons on the hydrodynamics of the 
interaction, the burst light-curve, its spectral hardness and softening. For the
fireball we first consider the impulsive model and after that we use the wind 
model to obtain a wider and less dense fireball when deceleration becomes 
important ($t_{dec}$) and thus a more relativistic reverse shock.

The peak of the synchrotron emission from the forward shock, which dominates the 
overall emission of the burst and the afterglow, is at an observer energy
\begin{equation}
h\nu_p = 1.4\times 10^{-8}\,(1+z)^{-1} \gamma_m^2 B' \Gamma\; {\rm eV} 
       = 10^{-3}\,[(1+z)/2]^{-1} \varepsilon_{el}^2 (\varepsilon_{mag} n_0)^{1/2} 
        \Gamma^4 \; {\rm eV} \;,
\label{epeak}
\end{equation}
where we used the synchrotron critical frequency corresponding to $\gamma_m$, 
averaged over the pitch angle,
and $n_0$ is the external medium particle density in ${\rm cm}^{-3}$.
For a fireball initial Lorentz factor $\Gamma_0$, the average flow Lorentz factor 
during the main burst is $\Gamma_{\gamma} \sim (1/2) \Gamma_0$.
In all simulations discussed below, the initial Lorentz factor is $\Gamma_0=500$ 
and the energy release parameters  $\varepsilon_{el}$ and $\varepsilon_{mag}$
have been chosen at the maximum value (\ie equipartition). Then equation 
(\ref{epeak}) gives $h\nu_p \sim 250 \,{\rm keV}$. For this $\Gamma_0$ values 
of these parameters too much below equipartition would lead to spectral peaks 
below the first BATSE channel (lower edge at $\sim 25 \,{\rm keV}$).
If the electron acceleration timescale is taken to be its gyration period, 
then $\gamma_M \sim 5 \times 10^7\, (B/{\rm 1\,G})^{-1/2} \sim 8 \times 10^7 \,
(\varepsilon_{mag} n_0)^{-1/4} \Gamma^{-1/2}$. Thus, during the GRB,
$\gamma_M/\gamma_m \sim 30\;\varepsilon_{el}^{-1} (\varepsilon_{mag} n_0)^{-1/4}
(\Gamma_0/500)^{-3/2}$, implying $\gamma_M/\gamma_m \sim 120$ for $\Gamma_0=500$
and equipartition. 

Figure 1 shows the average burst spectrum for an impulsive fireball with weak 
coupling in the shocked fluid. It has six components: one synchrotron and two 
inverse Compton from each shock. Behind the reverse shock, inverse Compton 
scatterings of locally produced photons take place in the Thomson regime; 
behind the forward shock these scatterings occur in the extreme Klein-Nishina 
regime and substantial up-scattered radiation is emitted only after electrons 
have cooled; mixed scatterings take place in a mild Klein-Nishina regime.
The steep cut-off that can be seen above the synchrotron FS peak in Figure 1 
(and also in Figure 2) is due to the fact that very energetic electrons 
($\gamma_M/\gamma_m >10$) have very short cooling
timescales and tracking their evolution accurately, by choosing a computational 
timestep smaller than this short synchrotron cooling timescale, would lead to
very long runs. Therefore the cut-off seen above $\sim 10$ MeV arises from 
choosing a low value of the ratio $\gamma_M/\gamma_m$ and should be regarded
as a deficiency of purely computational origin, and not as a
deficiency of the fireball model. 

In the weak coupling model, electrons are accelerated and exchange energy with 
protons and magnetic fields only at the shock, but not anywhere else in the
flow. Such electrons cool very 
fast in a burst that has $h\nu_p$ above 50 keV. Thus, on 
timescales shorter than $t_{dec}$, a good fraction of the available internal 
energy of the shocked fluid remains locked up in protons. This fraction is
larger than 1/2 because the fluid is continuously decelerated and more and 
more internal energy is produced. Nevertheless the evolution of the shocked 
fluid after electron cooling is not totally adiabatic, as the internal energy is 
used to drive forward the blast wave that sweeps up the external medium and to 
accelerate new electrons capable of radiating away the internal energy. 
On the other hand, if electrons are re-accelerated behind the forward shock 
(strong coupling model), then the internal energy is depleted very fast and 
the shocked structure stays in a radiative regime for longer times, until 
the electron themselves become adiabatic. If the re-acceleration timescale 
is much longer than the cooling timescale (as is the case in the weak coupling 
scenario), the electron spectrum will have an index $p/2$, due to the cooling 
and the continuous 
electron injection at the shock, while if the re-acceleration takes place on a
timescale shorter than the cooling timescale (weak coupling model), the spectral 
index will be $(1/2)(p-1)$. Therefore, one expects the strong coupling model 
burst to show a photon spectrum harder than for the weak coupling model. 
The self-inverse Compton scatterings are in the extreme Klein-Nishina regime 
for a longer time than in the weak coupling model, making the inverse Compton 
scattering less efficient for electron cooling (see Figure 2). 

In the {\it impulsive model}, the co-moving frame density of the fireball when 
deceleration becomes important (at time $t \siml t_{dec}$) is $\sim \Gamma_0^2$ 
times larger than that of the external medium. The reverse shock is 
quasi-newtonian ($\Gamma_R \simeq 1.1$) in the frame of the yet unshocked 
fireball, while the forward shock moves in the lab-frame with $\Gamma_F \simg 
300$. Such a reverse shock is inefficient in converting the ejecta's kinetic 
energy into heat (the ratio of the internal and rest-mass energy density behind 
the shock is $\Gamma_R -1$). A more relativistic reverse shock can be obtained 
if the fireball is less dense, which, for the same mass, requires a larger 
thickness. This can be achieved if the fireball results from an energy release 
that lasted more than few seconds ({\it wind model}). For a wind of duration 
$t_{wind} = 33$ s, the lab-frame fireball thickness is $10^{12}$ cm, $\sim 10$ 
times larger than in the impulsive scenario. In this case $\Gamma_R \simeq 
1.5$ and $\Gamma_F \simeq 180$.  A more relativistic reverse shock radiates 
more efficiently, while a less relativistic forward shock leads to a softer 
and weaker GRB, as shown in Figure 2 (long dashed curve) by the relative intensity 
and position of the peaks of the synchrotron emission from the two shocks.

In Figure 3 we compare the light-curves and spectral evolution of the bursts 
obtained in three models: impulsive with weak or strong coupling, and wind 
with strong coupling. The left graph legend indicates that the reverse shock 
contributes more to the observed burst if there is a strong coupling, which 
is due to the fact that the continuously generated internal energy behind 
the reverse shock is radiated, rather than being stored in protons and used 
to push the forward shock (and thus released in the end by the forward shock). 
The same legend gives the temporal asymmetry defined as $\int_{T_p}^{\infty} 
F_{23}(t) dT / \int_{0}^{T_p} F_{23}(T) dT$, where $F_{23}$ is the flux in 
the second and third BATSE channels (50 keV -- 1 MeV) and $T_p$ is the peak time.
The temporal asymmetry observed in real bursts is between 1.4 and 2.0 (Mitrofanov 
\etal 1996), smaller than that of the bursts arising from an impulsive release 
of the ejecta. The $T^{-\alpha}$ decay of the burst is steepest in the wind 
model (see legend of left graph in Figure 3). The burst $\gamma$-ray efficiency (ratio of 
fluence in the BATSE four channels, 25 keV -- 1 MeV, and the bolometric fluence) 
is the lowest in the same model, $\sim 30$\% compared to the 50\% efficiency 
reached in the impulsive models.  The right graph of Figure 3 shows that for 
the same parameters $(\Gamma_0,n_0;\varepsilon_{mag},\varepsilon_{el})$, 
the wind model produces 
a soft burst with the slowest softening rate, while the impulsive model with 
strong coupling gives the hardest burst. One would expect these features to 
be present not only in single-hump bursts but also in the individual
pulses of those bursts with a modest time variability. 

 The external shock model is able to explain well established features of
GRBs, such as (1) the brightness--spectral hardness correlation, (2) the spectral
hardening before an intensity pulse superposed on a continuous spectral 
softening, and (3) pulses peak earlier and last shorter at higher energies,
as shown in Panaitescu \& \Mesz (1998a). The synchrotron thin model
considered here is not able to account for values of the low energy spectral slope
larger than 1/3 (4/3 for $\nu F_{\nu}$) (Crider \etal 1997, Preece \etal 1998,
Strohmayer \etal 1998).
As mentioned in the previous section, the external shock model is not able 
to explain without the use of extra assumptions (\eg a rapidly varying 
magnetic field combined with a highly anisotropic radiation emission in the 
co-moving frame)
the large number of pulses observed in many GRBs. We have considered here GRB
light-curves and spectra in the framework of this model only for the purpose 
of comparing the effects of an extended and impulsive energy release and 
those of a 
strong and weak coupling in the fireball. Even if the burst itself arises
from internal shocks (Rees \& \Mesz 1994), one expects the external shock
to be at work when the fireball runs into the surrounding medium.

\section{Afterglows}

The spectral evolution of the afterglow is mainly determined by that of 
the bulk Lorentz factor of the shocked fluid and we will assume all other 
parameters (such as $\varepsilon_{mag}$ and $\varepsilon_{el}$) to be constant. 
We consider the impulsive model with 
weak or strong coupling in the remnant. In the absence of a strong
coupling, electrons cool fast and most of the burst emission comes from 
the leading edge of the shocked external matter (immediately behind the 
blast wave).  If a strong coupling is present, then all the fluid heated
by the two shocks radiates efficiently.

The general expected behavior of $\Gamma$ during the relativistic phase 
is $\propto t^{-n}$, where $t$ is the lab-frame time, $n=3$ if the remnant 
is radiative, and $n=3/2$ if it is adiabatic (Blandford \& McKee 1976). 
Numerically we found that the remnant Lorentz factor can be approximated by
\begin{equation}
\Gamma \simeq \Gamma_{\gamma} (t/t_{dec})^{-n} \;,
\label{Gamma} 
\end{equation}
for $t>t_{dec}$ and before the beginning of the non-relativistic regime,
where
\begin{equation}
t_{dec} = 1.4 \times 10^6\, (E_{52}/n_0)^{1/3} (\Gamma_0/500)^{-2/3}\, {\rm s}\; .
\label{tdec}
\end{equation}
$t_{dec}$ is the time when the mass of the swept up external fluid is a fraction 
$\Gamma_0^{-1}$ of the ejecta's mass, representing the deceleration onset time
(Rees \& \Mesz 1992). In equation (\ref{tdec}) 
$E = 10^{52} E_{52} \omega_{jet}$ ergs is the total amount of energy
released in the ejecta, where $\omega_{jet}$ is the solid angle of the ejecta. 
In a radiative remnant the electrons must be themselves radiative: 
$t_{sy} < t_{ad}$. Here $t_{sy}$ is the lab-frame electron cooling timescale, 
which is practically determined only by synchrotron cooling:
\begin{equation}
t_{sy}(\gamma_m)= 8.4 \times 10^6\, (\varepsilon_{el} \varepsilon_{mag} 
            n_0)^{-1} \Gamma^{-2}\; {\rm s}\;,
\label{tsy}
\end{equation}
and $t_{ad}$ is the adiabatic cooling timescale. The shell of shocked external 
matter is compressed between the contact discontinuity and the forward shock, the
increase in the shell thickness in time being rather due to the continuous accumulation
of external matter than to a radial expansion of the shell. Thus, to a good 
approximation, $t_{ad}=(2^{3/2}-1)\,t=1.83\,t$.
Using equations (\ref{Gamma}) and (\ref{tsy}), it results that for $n=3$ electrons 
become adiabatic when $\Gamma$ drops below
\begin{equation}
\Gamma_{r \rightarrow a} = 2.5\; (9\,\varepsilon_{el} \varepsilon_{mag})^{-3/5} 
          n_0^{-2/5} E_{52}^{-1/5} (\Gamma_0/500)^{1/5} \;.
\label{Gtrans}
\end{equation}
Therefore, at equipartition, electrons are radiative as long as the radiative 
remnant is relativistic. 

The evolution of Lorentz factor of the fluid moving on the line of sight
toward the center ($lsc$) of explosion (\ie pointing exactly toward the observer) 
can be calculated analytically from $dT=(1+z)\,dt/(4\,\Gamma^2)$, where $T$ is the
arrival time of the photons emitted at shock and on the $lsc$. 
The result can be cast into the simple form
\begin{equation}
 \Gamma_{lsc}(T)= C_n\, \Gamma_{\gamma} (T/T_{\gamma})^{-n/(2n+1)} \;,
\label{Glsc}
\end{equation}
where $T_{\gamma} \equiv 2\,(1+z)\,t_{dec} \Gamma_{\gamma}^{-2} = 92\; [(1+z)/2]
(E_{52}/n_0)^{1/3} \Gamma_0^{-8/3}\; {\rm s}$ is a good 
approximation\footnotemark of the $\gamma$-ray burst duration.
\footnotetext{This definition was chosen to match the duration obtained numerically
              (see Figure 3, left graph) and is larger by a factor 16 than the usual 
              result $T_{\gamma} \sim (1+z)\,t_{dec}/(2\,\Gamma_0^2)$, 
              which does not take into account the angular spreading 
              contribution to the burst duration, and the fact that 
              $\Gamma \siml \Gamma_0/2$ during most of the $\gamma$-ray emission} 
For a radiative remnant ($n=3$), one can show that $C_{3}=0.18$. If there is a strong
coupling, the remnant and the electrons become adiabatic simultaneously, at a 
time $T_{r \rightarrow a}$ that can be calculated using equations (\ref{Gtrans})
and (\ref{Glsc}).
After that, the evolution of $\Gamma_{lsc}$ is given by equation (\ref{Glsc}) 
with $n=3/2$ and a coefficient 
\begin{equation}
 C_{3/2} = 0.13\; (9\,\varepsilon_{el} 
 \varepsilon_{mag})^{-3/40} n_0^{-1/20} E_{52}^{-1/40} (\Gamma_0/500)^{-1/10} 
\label{C32}
\end{equation}
that has a very weak dependence on the burst parameters. 
For the weak coupling remnant we found numerically that (if the energy release
parameters are not much below equipartition) the quasi-adiabatic regime starts
early in the afterglow, at times when the spectrum peaks in the soft UV.
An analytic calculation of the time when the weak coupling remnant becomes
adiabatic is too inaccurate and we shall further use for the 
coefficient in equation (\ref{Glsc}) a value inferred from numerical results:
$C_{3/2}=0.32$. Thus, for the adiabatic remnant with weak coupling
\begin{equation}
\Gamma_{lsc,wc} =  6.1\; (E_{52}/n_0)^{1/8} [(1+z)/2]^{3/8} T_d^{-3/8} \;,
\label{Gwc}
\end{equation}
where $T_d$ is the observer time measured in days. 
Note that if $C_{3/2}$ does not depend too strong on the burst parameters 
(as suggested by the eq. [\ref{C32}] for the strong coupling case), 
then $\Gamma_{lsc,wc}$ has a weak dependence on the model parameters.
If the evolution of $\Gamma$ is the 
most important factor in determining the afterglow's features, 
then external shock GRBs arising from fireballs with different $\Gamma_0$'s,
exhibiting thus very different timescales,
should be followed by afterglows that have similar timescales. 

The afterglow that follows the burst of Figure 1 (weak coupling) is shown in 
Figure 4.  As the forward shock decelerates, the synchrotron emission from it 
shifts toward lower energies as $h\nu_p \propto T^{-1.4}$, consistent with 
the adiabatic regime of the remnant $\Gamma \propto t^{-1.5}$. At all times 
the intensity of the IC up-scattered emission is below that of the synchrotron 
one, which shows that inverse Compton is less efficient in electron cooling 
than synchrotron emission.  90\% of the initial fireball energy is released
during the weak-coupling afterglow shown in Figure 4.
The wide-band distribution of the energy radiated is: 35\% as $\gamma$-rays 
(above 100 keV), 35\% as X-rays (1 keV -- 100 keV), 21\% in the UV (1 eV -- 1 keV),
5\% in optical (1 eV -- 10 eV) and 5\% in IR and radio (below 1 eV). 
For a strong coupling remnant, the distribution is 49\%, 22\%, 19\%, 5\% 
and 5\%, respectively. Note that the strong coupling case leads to a higher 
$\gamma$-ray fluence at the expense of a lower X-ray fluence. 

Equation (\ref{epeak}) gives the peak of the synchrotron radiation assuming 
that the $\gamma_m$-electrons give almost all the burst radiation, which is 
correct only if electrons are not re-accelerated on a very short timescale. 
This equation may under-estimate the true value of $h\nu_p$ in the case of 
a continuous post-shock re-acceleration on a timescale shorter
than the cooling timescale, when $\nu F_{\nu}$ is expected to have a positive 
slope $(1/2)(3-p)$ for $p < 3$. In this case,
due the fact that photons received simultaneously by the observer were emitted
at different lab-frame times, thus from shocked material with different $\gamma_m$,
$B'$ and $\Gamma$, the real peak of $\nu F_{\nu}$ is at a frequency higher
than $\nu(\gamma_m)$, that cannot be accurately calculated analytically. For
simplicity, we will proceed with the analytical derivations assuming 
that the synchrotron spectrum peak $\nu_p$ is determined only by $\gamma_m$, 
as given by equation (\ref{epeak}).

An estimate of the time $T_{\nu_p}$ when the peak of $\nu F_{\nu}$  
reaches a given observational frequency $\nu_p$ and of the source size at that 
time can be obtained using the geometry of the equal arrival time surface
described by Panaitescu \& \Mesz (1998b). From equation (\ref{epeak}), the 
flow Lorentz factor of the fluid that gives most radiation at detector 
frequency $\nu_p$ is $\Gamma_{\nu_p}=(\nu_p/\nu_{\gamma})^{1/4}\Gamma_{\gamma}$, 
where $\nu_{\gamma}$ is the synchrotron peak frequency during the 
$\gamma$-ray burst. Most of this fluid is off-set from the $lsc$
and we shall denote by $f_{\|}$ the ratio between the projection onto
the $lsc$ of the radial coordinate (measured from the center of explosion) 
of the region that gives most of the radiation and the radial coordinate of 
the fluid on the $lsc$. This ratio must be determined from the geometry 
of the equal arrival time surface.
The Lorentz factor of the fluid on the $lsc$ and on the equal-$T_{\nu_p}$ 
surface is $\Gamma_{lsc} = f_{\|}^n \Gamma_{\nu_p}=f_{\|}^n 
(\nu_p/\nu_{\gamma})^{1/4} \Gamma_{\gamma}$. Using equation (\ref{Glsc}),
$T_{\nu_p} = (C_n^{1/n} f_{\|}^{-1})^{2n+1}
(\nu_{\gamma}/\nu_p)^{(2n+1)/4n} T_{\gamma}$.
For an adiabatic remnant ($n=3/2$, $C_{3/2}=0.32$, $f_{\|}=0.82$)
\begin{equation}
T_{\nu_p} = 6.3\, \left( \frac{h\nu_{\gamma}}{100\,{\rm keV}} \right)^{2/3} 
 \left( \frac{h\nu_p}{1\,{\rm eV}} \right)^{-2/3} T_{\gamma,2}\; {\rm hours} \;,
\label{time}
\end{equation}
where $T_{\gamma}=100\,T_{\gamma,2}\;{\rm s}$. 
For the afterglow shown in Figure 4, $h\nu_{\gamma}=250\;{\rm keV}$. 
Equation (\ref{time}) leads to $T_{1\,eV} = 11\,{\rm h}$, consistent with 
the afterglow spectral softening shown in Figure 4. 

The size of the source at time $T$, as seen  
projected on a plane perpendicular to the $lsc$,
is $R_{\bot} = f_{\bot}\Gamma_{lsc} [cT/(1+z)]$, where $f_{\bot} = 2^{3/2} (2n+1)
[2(n+1)]^{-(n+1)/(2n+1)}$. Using (\ref{Glsc}) with $\Gamma_{\gamma}$ expressed
as a function of $\nu_{\gamma}$ with the aid of equation (\ref{epeak}), 
one finds for an adiabatic remnant ($n=3/2$, $f_{\bot}=4.1$) that, at equipartition,
\begin{equation}
 R_{\bot} = 2.7 \times 10^{16}\, \left( \frac{1+z}{2} \right)^{-3/4} n_0^{-1/8}
 \left( \frac{h\nu_{\gamma}}{100\,{\rm keV}} \right)^{1/4}
 T_{\gamma,2}^{3/8} T_d^{5/8} \; {\rm cm}\; .
\label{size}
\end{equation}
Equation (\ref{size}) relates the remnant size to characteristics of the main
burst. Using equation (\ref{Gwc}), the same size can be written as:
\begin{equation}
 R_{\bot} = 3.2 \times 10^{16}\,(E_{52}/n_0)^{1/8} [(1+z)/2]^{-5/8} T_d^{5/8} \; {\rm cm}\; .
\end{equation}
For the afterglow shown in Figure 4, $R_{\bot} = 3.2 \times 10^{16}\,
T_d^{5/8}\; {\rm cm}$, thus the apparent source radius evolves as
$\phi = 1.8\; T_d^{5/8}\,{\rm \mu as}$.
The source appears to the observer as a disk that is brighter near the edge
than near the center. The width of the outer ring that radiates 50\% of the
radiation is $\sim 0.19\,\phi$. 
Equations (\ref{time}) and (\ref{size}) can be used to test the fireball model, 
once the duration and peak frequency of the main burst are measured. 

Figure 5 shows a comparison between the time histories of the two bursts that 
had afterglows below the X-ray domain (February 28 and May 08) and the numerical 
simulations in both strong and weak coupling models.  There are several 
disagreements between the real afterglows and the simulated ones, \eg the much 
lower fluence of the February 28 burst in the $0.1\,{\rm keV} - 2\,{\rm keV}$ 
band at $T < 1$ day, the absence of a rise in the numerical V magnitude and 
the larger numerical flux densities at 4.9 GHz. The relatively steep falls that 
can be seen at $T \simg 1$ day are due to the computational constraint described 
in the previous section, regarding the very short cooling synchrotron cooling 
time of the electrons in the high energy part of the power-law distribution, 
\ie those electrons which give the X-ray emission.

The optical and radio fluxes we obtained may be higher than what was observed 
for the afterglows of GRB 970228 and 090508 due to the large fireball energy 
used in the numerical simulation\footnotemark.  
\footnotetext{A value of $E \sim 10^{53} \;{\rm ergs}$ is used for an isotropic
  fireball to simulate a burst with a peak flux in the BATSE window 
  $\sim 10^{-6}\;{\rm ergs\, cm^{-2} s^{-1}}$, corresponding to a peak photon 
  flux $\sim 1\;{\rm \gamma\, cm^{-2} s^{-1}}$, given a smooth external shock 
  light-curve (multi-peaked bursts can reach a higher peak flux with a lower 
  energy budget). Shorter duration bursts, with higher peak fluxes and requiring 
  less energy, can be simulated using larger Lorentz factors $\Gamma_0$, which 
  would however require longer numerical runs.}
The dependence on the burst parameters of the flux received at a fixed  
frequency $\nu$ can be obtained analytically from the flux at the peak frequency 
$F_{\nu_p} = (\Gamma T)^2 (\Gamma^3 I'_{\nu'_p})$ (\Mesz \& Rees 1997),
where $I'_{\nu'_p} \sim (n'_e/4\,\pi) (P'_{sy}/\nu'_p) \min{(ct'_{sy},\Delta')}$ 
is the co-moving intensity at the peak of the synchrotron
emission from the least energetic electrons ($\gamma_m$). 
Here $n'_e$, $P'_{sy}$, $t'_{sy}$ and $\Delta'$ 
are the co-moving electron density, synchrotron power, cooling timescale and 
remnant thickness. These quantities can be easily calculated and put together
to yield $F_{\nu_p}$ as a function of $\Gamma \propto \Gamma_{lsc}$, 
which is given by equation (\ref{Gwc}) for an adiabatic remnant with weak coupling.
One finds that for $\nu < \nu_p$
$F_{\nu} = (\nu/\nu_p)^{1/3} F_{\nu_p} \propto \varepsilon_{mag}^{-2/3} 
\varepsilon_{el}^{-5/3} E^{1/3} T \nu^{1/3}$ if electrons are radiative, and  
$F_{\nu} \propto \varepsilon_{mag}^{1/3} \varepsilon_{el}^{-2/3} n_0^{1/2} 
E^{5/6} T^{1/2} \nu^{1/3}$ if electrons are adiabatic, assuming in both cases 
an adiabatic remnant. Note that in the latter case $F_{\nu}$ depends rather 
strongly on the burst energy. 
At frequencies $\nu > \nu_p$ the monochromatic flux is
$F_{\nu} = (\nu/\nu_p)^{-p/2} F_{\nu_p} \propto 
\varepsilon_{mag}^{1/8} \varepsilon_{el}^{3/2} E^{9/8} T^{-11/8} \nu^{-5/4}$ 
if electrons are radiative, and $F_{\nu} = (\nu/\nu_p)^{-(p-1)/2} F_{\nu_p}
\propto \varepsilon_{mag}^{7/8} 
\varepsilon_{el}^{3/2} n_0^{1/2} E^{11/8} T^{-9/8} \nu^{-3/4}$ if electrons 
are adiabatic, for $p=2.5$ . In both cases, $F_{\nu}$ has a strong dependence
on $\varepsilon_{el}$ and the available energy $E$. The previous relationships 
were derived assuming that all electrons are in the same radiative regime. In 
reality, high energy electrons can remain radiative for much longer times than 
those with $\gamma_m$, altering somewhat the power-law indices derived above. 
For example, if $\gamma_m$-electrons become adiabatic at time $T_m $,
electrons with $\gamma_e = 2 \gamma_m$ become 
adiabatic when $\Gamma$ has decreased by a factor $\sqrt{2}$, which occurs at
time $T_e = (\sqrt{2})^{8/3} T_m = 2.5\, T_m$.

Over-estimations of the radio and optical fluxes may also arise from the fact that
we assumed a spherically symmetric ejecta. So far we worked with an initial energy 
release of $10^{52}$ ergs/sr, which is consistent with the ``traditional" release of 
$10^{51}$ ergs if the ejecta are collimated in a jet of solid angle 0.1 sr 
(half angular opening $\simeq 11^{\rm o}$). Thus the observer would see the edge of 
this jet when $\Gamma$ drops below 5, corresponding to $T \simg$ 2 day in our 
example. The error made in the numerical simulation, which assumed spherical 
symmetry, increases with time as the jet opening angle becomes smaller and 
smaller than $\Gamma^{-1}$, the angular opening of the surface over which the 
emitted radiation was integrated. Thus fluxes are over-estimated by a factor 
$\sim 1.2$ (approximately 0.2 magnitudes) at $T=2$ days and by a factor $2.2$ 
(corresponding to 0.8 magnitudes) at $T=10$ days.

Numerical over-estimations of the flux at radio frequencies are also due to
the fact that we did not take into account the synchrotron self-absorption. 
It is easy to find out the time dependence of the
self-absorption frequency $\nu_{ab}$, but it is not a trivial exercise to calculate 
an accurate value of it. If the synchrotron self-absorption coefficient is used,
one should take into account the relativistic expansion of the absorbing fluid 
during the propagation of a photon through it. Here, we shall use the equality of the
self-absorbed intensity $(I'_{\nu'})_{abs} \sim 2\,\gamma_m m_e \nu'^2$ 
($m_e$ is the electron's mass) and the synchrotron intensity
$(I'_{\nu'})_{p-law} \sim (\nu'/\nu'_p)^{1/3} I'_{\nu'_p}$ for a power-law distribution 
of electrons, at the unknown frequency $\nu'_{ab}$ (primed quantities are measured
in the co-moving frame). In both kinds of calculations of $\nu_{ab}$, 
the shape of the equal arrival time surface must be taken into account.
In the method chosen here to calculate $\nu_{ab}$, this is taken into account
by using the Lorentz factor $\Gamma$ of the fluid that dominates the flux received
by the observer, which is related to $\Gamma_{lsc}$ by $\Gamma = f_{\|}^{-n} \Gamma_{lsc}$.
The end result is that for an adiabatic remnant with 
weak coupling $\nu_{ab} = 0.37\; \varepsilon_{mag}^{-2/5} \varepsilon_{el}^{-8/5}
n_0^{3/10} E_{52}^{-1/10} [(z+1)/2]^{-13/10} T_d^{3/10}$ GHz if electrons are 
radiative, and $\nu_{ab} =  1.6\; \varepsilon_{mag}^{1/5} \varepsilon_{el}^{-1} 
n_0^{3/5} E_{52}^{1/5} [(z+1)/2]^{-1} T_d^0$ GHz if electrons are adiabatic. 
For the representative values used so far and at equipartition, one obtains
$\nu_{ab} = 3.3\; T_d^{3/10}$ GHz for radiative electrons and $\nu_{ab} = 3.8\; 
T_d^0$ GHz for adiabatic electrons.  Thus one expects the remnant to be  
optically thin at 8.46 GHz, $\tau_{\rm 4.86\;GHz}\sim 1$, and optically thick 
at 1.43 GHz, consistent with the radio observations of the GRB 970508's 
afterglow (Frail 1997). Note that an electron parameter $\varepsilon_{el}$
below equipartition leads to higher optical depths at 4.9 GHz and that $\nu_{ab}$
depends weaker on the other model parameters. 

If there is a flat, low energy tail of electrons factor below $\gamma_m$, so that
$dN_e/d\gamma_e \propto \gamma_e^0$ as considered
by Vietri (1997) and Waxman (1997a), then the self-absorption frequency is 
$\nu_{ab} = 2.8\; \varepsilon_{mag}^{-1/4} \varepsilon_{el}^{-1} n_0^{1/4} 
[(z+1)/2]^{-1} T_d^0$ GHz if electrons are radiative, and $\nu_{ab} = 9.3\;\, 
\varepsilon_{mag}^{1/4} \varepsilon_{el}^{-1/2} n_0^{1/2} E_{52}^{1/4} 
[(z+1)/2]^{-3/4} T_d^{-1/4}$ GHz for adiabatic electrons. With the usual 
parameters, $\nu_{ab} = 11\; T_d^0$ GHz for radiative electrons and $\nu_{ab} = 12\; 
T_d^{-1/4}$ GHz for adiabatic electrons, implying optical thickness at  
4.86 GHz for $T < 37$ days. However, the remnant optical thickness at this frequency can 
be substantially lower if the ejecta was initially beamed and if, at low bulk 
Lorentz factors, the shocked fluid expands outside the cone, leading to a 
stronger deceleration than predicted by equation (\ref{Glsc}). If the effect of 
this sideways escape of the fluid is parameterized by including an extra 
factor $-A$ (with $A > 0$) in the exponent of $T$ in equation (\ref{Glsc}),
then $\nu_{ab} = 12\; T_d^{-(A+1/4)}$ GHz for adiabatic electrons, yielding 
optical thinness at 4.86 GHz at times $T > 5$ days if $A > 1/3$. Such a power-law
approximation is suitable only for short times, as the decay of $\Gamma$ due
to the sideways escape is in fact exponential (Rhoads 1997): $\Gamma \propto
\exp(-t/t_{se})$, where $t_{se} = (2\,\kappa \Gamma_0^2 \theta_{jet}^2)^{1/3}
\,t_{dec} = 2.5\; (\Gamma_0/500)^{2/3} (\theta_{jet}/1^{\rm o})^{2/3} 
(\kappa/0.1)^{1/3}\,t_{dec}$, $\kappa$ being the fraction of the initial 
energy contained in the remnant at the onset of the adiabatic phase 
(assumed to start before the effect of the sideways expansion becomes important) 
and $\theta_{jet}$ being the half-angular opening of the jet.

Throughout this work it was assumed that the external medium is homogeneous.
For a medium varying as a power law in the distance and considering 
only the case of a relativistic remnant
and a power-law spectrum $F_{\nu} = (\nu/\nu_p)^a F_{\nu_p}$, one finds that
$F_{\nu} \propto T^{[4-\alpha+a(24-7\alpha)]/(14-4\alpha)}$ for a radiative 
remnant, $F_{\nu} \propto T^{(3a+1)/2}$ for radiative electrons in an adiabatic 
remnant and $F_{\nu} \propto T^{(3a/2)-[\alpha/(8-2\alpha)]}$ for adiabatic 
electrons, where $\alpha < 3$ is the index of the external fluid density:
$n_{ext} \propto r^{-\alpha}$.  If the external medium index changes from 
$\alpha=0$ (homogeneous medium) to $\alpha=2$ (pre-ejected wind), then
the exponent of $T$ in the previous expressions for the evolution of $F_{\nu}$ 
changes from $(2/7)(6a+1)$ to $(1/3)(5a+1)$ for a radiative remnant, is constant 
for an adiabatic remnant with radiative electrons, and varies from $3a/2$ to 
$(3a-1)/2$ for an adiabatic remnant. Therefore, the slope of the decay of 
$F_{\nu}$ is altered significantly by the external medium density index only if 
the electrons are adiabatic. The size of the adiabatic remnant evolves as 
$R_{\bot} \propto T^{(5-\alpha)/(8-2\alpha)}$, which gives $R_{\bot} \propto T^{5/8}$ for
$\alpha=0$ and $R_{\bot} \propto T^{3/4}$ for $\alpha=2$. Thus, at given observer time,
the remnant interacting with a pre-ejected wind appears larger than one 
running into a homogeneous external medium. When electrons are adiabatic, the 
self-absorption frequency evolves as $\nu_{ab} \propto T^{-3\alpha/(20-5\alpha)}$,
which shows that, while $\nu_{ab}$ is constant in time for a homogeneous external
medium, it decreases in time as $T^{-3/5}$ for a pre-ejected wind.
It can be shown that electrons become adiabatic earlier in a pre-ejected wind 
and that the time when the remnant becomes non-relativistic increases strongly
with the external medium density index, the evolution remaining relativistic 
up to tens of years in the pre-ejected wind case.

In the analytic derivations above we made several 
assumptions regarding the isotropy of the ejecta, the electron radiative 
regimes and the viewing geometry, which could lead to substantial inaccuracies 
in the analytical power-laws describing the evolution of $F_{\nu}$, that are
used in comparisons between observations and predictions of the fireball model:
(1) the ejecta are isotropic (see \Mesz \etal 1998 for the wide 
    variety of slopes that can be obtained in non-isotropic models); 
(2) all electrons are assumed in the same radiative regime as those with the 
    minimum Lorentz factor (which leads to shallower spectral slopes at high 
    frequency), although the times when electrons with different $\gamma_e$ 
    become adiabatic may span more than two orders of magnitude; 
(3) at fixed time, the observer receives radiation emitted at a unique 
    lab-frame time (photons arriving simultaneously at detector may 
    have been emitted by gas shocked at different Lorentz factors; such mixing
    of radiation leading to a weaker spectral evolution of the afterglow); 
(4) the reverse shock emission can be neglected (low frequency radiation received 
    from this shock leads to shallower spectra in the infrared and to a continuous 
    decay of the radio fluxes, hindering the rise of the forward shock emission 
    in the early afterglow); 
(5) the model parameters for minimum electron Lorentz 
    factor and magnetic field strength are constant in time (parameters decaying as
    a power-law yield steeper rises of the low energy light-curves and steeper
    decays of the fluxes at high frequency). 
Approximations (1) and (5) are also made in the numerical calculations.  
Further complications may arise if the ejecta is beamed in a jet (Rhoads 1997) 
or if the external medium is inhomogeneous.

\section{Discussion}

Summarizing, the features of the bursts and afterglows arising from impulsive 
or wind fireballs with a strong or weak coupling of electrons with baryons and 
magnetic field, are: \\
\hspace*{3mm} 1. Wind fireballs produce softer GRBs than impulsive ones for the 
  same set of hydrodynamical and energy release parameters, and increase the 
  efficiency of the reverse shock, yielding a brighter optical and UV 
  counterpart. An appropriate change in model 
  parameters (particularly $\Gamma_0$) could shift the softer spectrum produced 
  by wind fireballs into the $\gamma$-ray domain, but the fact that the GRB 
  efficiency is lower than in the impulsive model makes this possibility
  less likely to be a real scenario. \\
\hspace*{3mm} 2. Strong coupling in the post-shock fluid leads to harder spectra 
  and to a radiative phase of the afterglow which extends to later times 
  than in the weak coupling case. It could explain the X-ray paucity observed in 
  many GRBs, by maintaining for longer times higher electron Lorentz factor and, 
  implicitly, the synchrotron emission from the blast wave in the $\gamma$-ray 
  range. \\
\hspace*{3mm} 3. At the same observer time, the intensity of the emission of a strong 
  coupling remnant at lower energies (radio) 
  is considerably weaker than that of a weak coupling one. In our simulations, 
  an isotropic fireball
  with strong coupling requires $10^{53}$ ergs to yield radio fluxes comparable to those
  observed in the afterglow of GRB 970508, which is one order of magnitude larger
  than that necessary for a weak coupling fireball. \\
\hspace*{3mm} 4. Unless the initial fireball Lorentz factor is substantially higher
  than considered here, the
  parameters describing the magnetic field strength and the electron energy 
  must not be too much below equipartition, otherwise the main burst would have 
  a spectral peak below $\sim 50$ keV. As a consequence of this, the remnant is not   
  adiabatic in the early afterglow. The adiabatic phase starts earlier for a 
  weak coupling remnant than for one with strong coupling. \\
\hspace*{3mm} 5. During the afterglow, the flow Lorentz factor of the shocked 
  fluid has only a weak dependence on the initial burst parameters, including the 
  fireball initial Lorentz factor, so that GRBs with very different peak fluxes
  and $\gamma$-ray durations, arising from fireballs interacting with homogeneous
  surrounding media, should be followed by afterglows whose time-scales are
  similar.

The results of the numerical hydrodynamic calculations presented here are 
meant to be illustrative, and are not intended as fits, but rather 
as a study of the observable consequences of various physical assumptions
about the energetic and dynamics which are possible in realistic models. 
An exploration of the parameter space will be needed, as well as consideration 
of some effects not included in the simplified model used here, such as
anisotropic distribution of energy in the ejecta, inhomogeneous external medium, 
time-varying energy release parameters, or jet-like ejecta, in order to 
provide a more detailed characterization of the external shock model and 
comparison with the observational material on bursts and their afterglows.

This research has been supported by NASA NAG5-2857 and NAG5-2362.

\clearpage

\def\PsfigVersion{1.9}
\ifx\undefined\psfig\else \fi

%

\let\LaTeXAtSign=\@
\let\@=\relax
\edef\psfigRestoreAt{\catcode`\@=\number\catcode`@\relax}
\catcode`\@=11\relax
\newwrite\@unused
\def\ps@typeout#1{{\let\protect\string\immediate\write\@unused{#1}}}
\ps@typeout{psfig/tex \PsfigVersion}


\def\figurepath{./}
\def\psfigurepath#1{\edef\figurepath{#1}}

%
%
\def\@nnil{\@nil}
\def\@empty{}
\def\@psdonoop#1\@@#2#3{}
\def\@psdo#1:=#2\do#3{\edef\@psdotmp{#2}\ifx\@psdotmp\@empty \else
    \expandafter\@psdoloop#2,\@nil,\@nil\@@#1{#3}\fi}
\def\@psdoloop#1,#2,#3\@@#4#5{\def#4{#1}\ifx #4\@nnil \else
       #5\def#4{#2}\ifx #4\@nnil \else#5\@ipsdoloop #3\@@#4{#5}\fi\fi}
\def\@ipsdoloop#1,#2\@@#3#4{\def#3{#1}\ifx #3\@nnil 
       \let\@nextwhile=\@psdonoop \else
      #4\relax\let\@nextwhile=\@ipsdoloop\fi\@nextwhile#2\@@#3{#4}}
\def\@tpsdo#1:=#2\do#3{\xdef\@psdotmp{#2}\ifx\@psdotmp\@empty \else
    \@tpsdoloop#2\@nil\@nil\@@#1{#3}\fi}
\def\@tpsdoloop#1#2\@@#3#4{\def#3{#1}\ifx #3\@nnil 
       \let\@nextwhile=\@psdonoop \else
      #4\relax\let\@nextwhile=\@tpsdoloop\fi\@nextwhile#2\@@#3{#4}}
%
\ifx\undefined\fbox
\newdimen\fboxrule
\newdimen\fboxsep
\newdimen\ps@tempdima
\newbox\ps@tempboxa
\fboxsep = 3pt
\fboxrule = .4pt
\long\def\fbox#1{\leavevmode\setbox\ps@tempboxa\hbox{#1}\ps@tempdima\fboxrule
    \advance\ps@tempdima \fboxsep \advance\ps@tempdima \dp\ps@tempboxa
   \hbox{\lower \ps@tempdima\hbox
  {\vbox{\hrule height \fboxrule
          \hbox{\vrule width \fboxrule \hskip\fboxsep
          \vbox{\vskip\fboxsep \box\ps@tempboxa\vskip\fboxsep}\hskip 
                 \fboxsep\vrule width \fboxrule}
                 \hrule height \fboxrule}}}}
\fi
%
%
\newread\ps@stream
\newif\ifnot@eof       
\newif\if@noisy        
\newif\if@atend        
\newif\if@psfile       
%
%
{\catcode`\%=12\global\gdef\epsf@start{
\def\epsf@PS{PS}
\def\epsf@getbb#1{%
%
%
\openin\ps@stream=#1
\ifeof\ps@stream\ps@typeout{Error, File #1 not found}\else
%
%
   {\not@eoftrue \chardef\other=12
    \def\do##1{\catcode`##1=\other}\dospecials \catcode`\ =10
    \loop
       \if@psfile
	  \read\ps@stream to \epsf@fileline
       \else{
	  \obeyspaces
          \read\ps@stream to \epsf@tmp\global\let\epsf@fileline\epsf@tmp}
       \fi
       \ifeof\ps@stream\not@eoffalse\else
%
%
       \if@psfile\else
       \expandafter\epsf@test\epsf@fileline:. \\%
       \fi
%
%
          \expandafter\epsf@aux\epsf@fileline:. \\%
       \fi
   \ifnot@eof\repeat
   }\closein\ps@stream\fi}%
%
%
\long\def\epsf@test#1#2#3:#4\\{\def\epsf@testit{#1#2}
			\ifx\epsf@testit\epsf@start\else
\ps@typeout{Warning! File does not start with `\epsf@start'.  It may not be a PostScript file.}
			\fi
			\@psfiletrue} 
%
%
{\catcode`\%=12\global\let\epsf@percent=
%
%
%
\long\def\epsf@aux#1#2:#3\\{\ifx#1\epsf@percent
   \def\epsf@testit{#2}\ifx\epsf@testit\epsf@bblit
	\@atendfalse
        \epsf@atend #3 . \\%
	\if@atend	
	   \if@verbose{
		\ps@typeout{psfig: found `(atend)'; continuing search}
	   }\fi
        \else
        \epsf@grab #3 . . . \\%
        \not@eoffalse
        \global\no@bbfalse
        \fi
   \fi\fi}%
%
%
\def\epsf@grab #1 #2 #3 #4 #5\\{%
   \global\def\epsf@llx{#1}\ifx\epsf@llx\empty
      \epsf@grab #2 #3 #4 #5 .\\\else
   \global\def\epsf@lly{#2}%
   \global\def\epsf@urx{#3}\global\def\epsf@ury{#4}\fi}%
%
%
\def\epsf@atendlit{(atend)} 
\def\epsf@atend #1 #2 #3\\{%
   \def\epsf@tmp{#1}\ifx\epsf@tmp\empty
      \epsf@atend #2 #3 .\\\else
   \ifx\epsf@tmp\epsf@atendlit\@atendtrue\fi\fi}


\chardef\psletter = 11 
\chardef\other = 12

\newif \ifdebug 
\newif\ifc@mpute 
\c@mputetrue 

\let\then = \relax
\def\r@dian{pt }
\let\r@dians = \r@dian
\let\dimensionless@nit = \r@dian
\let\dimensionless@nits = \dimensionless@nit
\def\internal@nit{sp }
\let\internal@nits = \internal@nit
\newif\ifstillc@nverging
\def \Mess@ge #1{\ifdebug \then \message {#1} \fi}

{ 
	\catcode `\@ = \psletter
	\gdef \nodimen {\expandafter \n@dimen \the \dimen}
	\gdef \term #1 #2 #3%
	       {\edef \t@ {\the #1}
		\edef \t@@ {\expandafter \n@dimen \the #2\r@dian}%
		\t@rm {\t@} {\t@@} {#3}%
	       }
	\gdef \t@rm #1 #2 #3%
	       {{%
		\count 0 = 0
		\dimen 0 = 1 \dimensionless@nit
		\dimen 2 = #2\relax
		\Mess@ge {Calculating term #1 of \nodimen 2}%
		\loop
		\ifnum	\count 0 < #1
		\then	\advance \count 0 by 1
			\Mess@ge {Iteration \the \count 0 \space}%
			\Multiply \dimen 0 by {\dimen 2}%
			\Mess@ge {After multiplication, term = \nodimen 0}%
			\Divide \dimen 0 by {\count 0}%
			\Mess@ge {After division, term = \nodimen 0}%
		\repeat
		\Mess@ge {Final value for term #1 of 
				\nodimen 2 \space is \nodimen 0}%
		\xdef \Term {#3 = \nodimen 0 \r@dians}%
		\aftergroup \Term
	       }}
	\catcode `\p = \other
	\catcode `\t = \other
	\gdef \n@dimen #1pt{#1} 
}

\def \Divide #1by #2{\divide #1 by #2} 

\def \Multiply #1by #2
       {{
	\count 0 = #1\relax
	\count 2 = #2\relax
	\count 4 = 65536
	\Mess@ge {Before scaling, count 0 = \the \count 0 \space and
			count 2 = \the \count 2}%
	\ifnum	\count 0 > 32767 
	\then	\divide \count 0 by 4
		\divide \count 4 by 4
	\else	\ifnum	\count 0 < -32767
		\then	\divide \count 0 by 4
			\divide \count 4 by 4
		\else
		\fi
	\fi
	\ifnum	\count 2 > 32767 
	\then	\divide \count 2 by 4
		\divide \count 4 by 4
	\else	\ifnum	\count 2 < -32767
		\then	\divide \count 2 by 4
			\divide \count 4 by 4
		\else
		\fi
	\fi
	\multiply \count 0 by \count 2
	\divide \count 0 by \count 4
	\xdef \product {#1 = \the \count 0 \internal@nits}%
	\aftergroup \product
       }}

\def\r@duce{\ifdim\dimen0 > 90\r@dian \then   
		\multiply\dimen0 by -1
		\advance\dimen0 by 180\r@dian
		\r@duce
	    \else \ifdim\dimen0 < -90\r@dian \then  
		\advance\dimen0 by 360\r@dian
		\r@duce
		\fi
	    \fi}

\def\Sine#1%
       {{%
	\dimen 0 = #1 \r@dian
	\r@duce
	\ifdim\dimen0 = -90\r@dian \then
	   \dimen4 = -1\r@dian
	   \c@mputefalse
	\fi
	\ifdim\dimen0 = 90\r@dian \then
	   \dimen4 = 1\r@dian
	   \c@mputefalse
	\fi
	\ifdim\dimen0 = 0\r@dian \then
	   \dimen4 = 0\r@dian
	   \c@mputefalse
	\fi
	\ifc@mpute \then
		\divide\dimen0 by 180
		\dimen0=3.141592654\dimen0
		\dimen 2 = 3.1415926535897963\r@dian 
		\divide\dimen 2 by 2 
		\Mess@ge {Sin: calculating Sin of \nodimen 0}%
		\count 0 = 1 
		\dimen 2 = 1 \r@dian 
		\dimen 4 = 0 \r@dian 
		\loop
			\ifnum	\dimen 2 = 0 
			\then	\stillc@nvergingfalse 
			\else	\stillc@nvergingtrue
			\fi
			\ifstillc@nverging 
			\then	\term {\count 0} {\dimen 0} {\dimen 2}%
				\advance \count 0 by 2
				\count 2 = \count 0
				\divide \count 2 by 2
				\ifodd	\count 2 
				\then	\advance \dimen 4 by \dimen 2
				\else	\advance \dimen 4 by -\dimen 2
				\fi
		\repeat
	\fi		
			\xdef \sine {\nodimen 4}%
       }}

\def\Cosine#1{\ifx\sine\UnDefined\edef\Savesine{\relax}\else
		             \edef\Savesine{\sine}\fi
	{\dimen0=#1\r@dian\advance\dimen0 by 90\r@dian
	 \Sine{\nodimen 0}
	 \xdef\cosine{\sine}
	 \xdef\sine{\Savesine}}}	      

\def\psdraft{
	\def\@psdraft{0}
}
\def\psfull{
	\def\@psdraft{100}
}

\psfull

\newif\if@scalefirst
\def\psscalefirst{\@scalefirsttrue}
\def\psrotatefirst{\@scalefirstfalse}
\psrotatefirst

\newif\if@draftbox
\def\psnodraftbox{
	\@draftboxfalse
}
\def\psdraftbox{
	\@draftboxtrue
}
\@draftboxtrue

\newif\if@prologfile
\newif\if@postlogfile
\def\pssilent{
	\@noisyfalse
}
\def\psnoisy{
	\@noisytrue
}
\psnoisy
\newif\if@bbllx
\newif\if@bblly
\newif\if@bburx
\newif\if@bbury
\newif\if@height
\newif\if@width
\newif\if@rheight
\newif\if@rwidth
\newif\if@angle
\newif\if@clip
\newif\if@verbose
\def\@p@@sclip#1{\@cliptrue}

\newif\if@decmpr


\def\@p@@sfigure#1{\def\@p@sfile{null}\def\@p@sbbfile{null}
	        \openin1=#1.bb
		\ifeof1\closein1
	        	\openin1=\figurepath#1.bb
			\ifeof1\closein1
			        \openin1=#1
				\ifeof1\closein1%
				       \openin1=\figurepath#1
					\ifeof1
					   \ps@typeout{Error, File #1 not found}
						\if@bbllx\if@bblly
				   		\if@bburx\if@bbury
			      				\def\@p@sfile{#1}%
			      				\def\@p@sbbfile{#1}%
							\@decmprfalse
				  	   	\fi\fi\fi\fi
					\else\closein1
				    		\def\@p@sfile{\figurepath#1}%
				    		\def\@p@sbbfile{\figurepath#1}%
						\@decmprfalse
	                       		\fi%
			 	\else\closein1%
					\def\@p@sfile{#1}
					\def\@p@sbbfile{#1}
					\@decmprfalse
			 	\fi
			\else
				\def\@p@sfile{\figurepath#1}
				\def\@p@sbbfile{\figurepath#1.bb}
				\@decmprtrue
			\fi
		\else
			\def\@p@sfile{#1}
			\def\@p@sbbfile{#1.bb}
			\@decmprtrue
		\fi}

\def\@p@@sfile#1{\@p@@sfigure{#1}}

\def\@p@@sbbllx#1{
		\@bbllxtrue
		\dimen100=#1
		\edef\@p@sbbllx{\number\dimen100}
}
\def\@p@@sbblly#1{
		\@bbllytrue
		\dimen100=#1
		\edef\@p@sbblly{\number\dimen100}
}
\def\@p@@sbburx#1{
		\@bburxtrue
		\dimen100=#1
		\edef\@p@sbburx{\number\dimen100}
}
\def\@p@@sbbury#1{
		\@bburytrue
		\dimen100=#1
		\edef\@p@sbbury{\number\dimen100}
}
\def\@p@@sheight#1{
		\@heighttrue
		\dimen100=#1
   		\edef\@p@sheight{\number\dimen100}
}
\def\@p@@swidth#1{
		\@widthtrue
		\dimen100=#1
		\edef\@p@swidth{\number\dimen100}
}
\def\@p@@srheight#1{
		\@rheighttrue
		\dimen100=#1
		\edef\@p@srheight{\number\dimen100}
}
\def\@p@@srwidth#1{
		\@rwidthtrue
		\dimen100=#1
		\edef\@p@srwidth{\number\dimen100}
}
\def\@p@@sangle#1{
		\@angletrue
		\edef\@p@sangle{#1} 
}
\def\@p@@ssilent#1{ 
		\@verbosefalse
}
\def\@p@@sprolog#1{\@prologfiletrue\def\@prologfileval{#1}}
\def\@p@@spostlog#1{\@postlogfiletrue\def\@postlogfileval{#1}}
\def\@cs@name#1{\csname #1\endcsname}
\def\@setparms#1=#2,{\@cs@name{@p@@s#1}{#2}}
%
%
\def\ps@init@parms{
		\@bbllxfalse \@bbllyfalse
		\@bburxfalse \@bburyfalse
		\@heightfalse \@widthfalse
		\@rheightfalse \@rwidthfalse
		\def\@p@sbbllx{}\def\@p@sbblly{}
		\def\@p@sbburx{}\def\@p@sbbury{}
		\def\@p@sheight{}\def\@p@swidth{}
		\def\@p@srheight{}\def\@p@srwidth{}
		\def\@p@sangle{0}
		\def\@p@sfile{} \def\@p@sbbfile{}
		\def\@p@scost{10}
		\def\@sc{}
		\@prologfilefalse
		\@postlogfilefalse
		\@clipfalse
		\if@noisy
			\@verbosetrue
		\else
			\@verbosefalse
		\fi
}
%
%
\def\parse@ps@parms#1{
	 	\@psdo\@psfiga:=#1\do
		   {\expandafter\@setparms\@psfiga,}}
%
%
\newif\ifno@bb
\def\bb@missing{
	\if@verbose{
		\ps@typeout{psfig: searching \@p@sbbfile \space  for bounding box}
	}\fi
	\no@bbtrue
	\epsf@getbb{\@p@sbbfile}
        \ifno@bb \else \bb@cull\epsf@llx\epsf@lly\epsf@urx\epsf@ury\fi
}	
\def\bb@cull#1#2#3#4{
	\dimen100=#1 bp\edef\@p@sbbllx{\number\dimen100}
	\dimen100=#2 bp\edef\@p@sbblly{\number\dimen100}
	\dimen100=#3 bp\edef\@p@sbburx{\number\dimen100}
	\dimen100=#4 bp\edef\@p@sbbury{\number\dimen100}
	\no@bbfalse
}
\newdimen\p@intvaluex
\newdimen\p@intvaluey
\def\rotate@#1#2{{\dimen0=#1 sp\dimen1=#2 sp
		  \global\p@intvaluex=\cosine\dimen0
		  \dimen3=\sine\dimen1
		  \global\advance\p@intvaluex by -\dimen3
		  \global\p@intvaluey=\sine\dimen0
		  \dimen3=\cosine\dimen1
		  \global\advance\p@intvaluey by \dimen3
		  }}
\def\compute@bb{
		\no@bbfalse
		\if@bbllx \else \no@bbtrue \fi
		\if@bblly \else \no@bbtrue \fi
		\if@bburx \else \no@bbtrue \fi
		\if@bbury \else \no@bbtrue \fi
		\ifno@bb \bb@missing \fi
		\ifno@bb \ps@typeout{FATAL ERROR: no bb supplied or found}
			\no-bb-error
		\fi
		%
%
		\count203=\@p@sbburx
		\count204=\@p@sbbury
		\advance\count203 by -\@p@sbbllx
		\advance\count204 by -\@p@sbblly
		\edef\ps@bbw{\number\count203}
		\edef\ps@bbh{\number\count204}
		\if@angle 
			\Sine{\@p@sangle}\Cosine{\@p@sangle}
	        	{\dimen100=\maxdimen\xdef\r@p@sbbllx{\number\dimen100}
					    \xdef\r@p@sbblly{\number\dimen100}
			                    \xdef\r@p@sbburx{-\number\dimen100}
					    \xdef\r@p@sbbury{-\number\dimen100}}
%
                        \def\minmaxtest{
			   \ifnum\number\p@intvaluex<\r@p@sbbllx
			      \xdef\r@p@sbbllx{\number\p@intvaluex}\fi
			   \ifnum\number\p@intvaluex>\r@p@sbburx
			      \xdef\r@p@sbburx{\number\p@intvaluex}\fi
			   \ifnum\number\p@intvaluey<\r@p@sbblly
			      \xdef\r@p@sbblly{\number\p@intvaluey}\fi
			   \ifnum\number\p@intvaluey>\r@p@sbbury
			      \xdef\r@p@sbbury{\number\p@intvaluey}\fi
			   }
			\rotate@{\@p@sbbllx}{\@p@sbblly}
			\minmaxtest
			\rotate@{\@p@sbbllx}{\@p@sbbury}
			\minmaxtest
			\rotate@{\@p@sbburx}{\@p@sbblly}
			\minmaxtest
			\rotate@{\@p@sbburx}{\@p@sbbury}
			\minmaxtest
			\edef\@p@sbbllx{\r@p@sbbllx}\edef\@p@sbblly{\r@p@sbblly}
			\edef\@p@sbburx{\r@p@sbburx}\edef\@p@sbbury{\r@p@sbbury}
		\fi
		\count203=\@p@sbburx
		\count204=\@p@sbbury
		\advance\count203 by -\@p@sbbllx
		\advance\count204 by -\@p@sbblly
		\edef\@bbw{\number\count203}
		\edef\@bbh{\number\count204}
}
%
%
\def\in@hundreds#1#2#3{\count240=#2 \count241=#3
		     \count100=\count240	
		     \divide\count100 by \count241
		     \count101=\count100
		     \multiply\count101 by \count241
		     \advance\count240 by -\count101
		     \multiply\count240 by 10
		     \count101=\count240	
		     \divide\count101 by \count241
		     \count102=\count101
		     \multiply\count102 by \count241
		     \advance\count240 by -\count102
		     \multiply\count240 by 10
		     \count102=\count240	
		     \divide\count102 by \count241
		     \count200=#1\count205=0
		     \count201=\count200
			\multiply\count201 by \count100
		 	\advance\count205 by \count201
		     \count201=\count200
			\divide\count201 by 10
			\multiply\count201 by \count101
			\advance\count205 by \count201
		     \count201=\count200
			\divide\count201 by 100
			\multiply\count201 by \count102
			\advance\count205 by \count201
		     \edef\@result{\number\count205}
}
\def\compute@wfromh{
		\in@hundreds{\@p@sheight}{\@bbw}{\@bbh}
		\edef\@p@swidth{\@result}
}
\def\compute@hfromw{
	        \in@hundreds{\@p@swidth}{\@bbh}{\@bbw}
		\edef\@p@sheight{\@result}
}
\def\compute@handw{
		\if@height 
			\if@width
			\else
				\compute@wfromh
			\fi
		\else 
			\if@width
				\compute@hfromw
			\else
				\edef\@p@sheight{\@bbh}
				\edef\@p@swidth{\@bbw}
			\fi
		\fi
}
\def\compute@resv{
		\if@rheight \else \edef\@p@srheight{\@p@sheight} \fi
		\if@rwidth \else \edef\@p@srwidth{\@p@swidth} \fi
}
%
\def\compute@sizes{
	\compute@bb
	\if@scalefirst\if@angle
	\if@width
	   \in@hundreds{\@p@swidth}{\@bbw}{\ps@bbw}
	   \edef\@p@swidth{\@result}
	\fi
	\if@height
	   \in@hundreds{\@p@sheight}{\@bbh}{\ps@bbh}
	   \edef\@p@sheight{\@result}
	\fi
	\fi\fi
	\compute@handw
	\compute@resv}

%
%
\def\psfig#1{\vbox {
	%
	\ps@init@parms
	\parse@ps@parms{#1}
	\compute@sizes
	\ifnum\@p@scost<\@psdraft{
		\special{ps::[begin] 	\@p@swidth \space \@p@sheight \space
				\@p@sbbllx \space \@p@sbblly \space
				\@p@sbburx \space \@p@sbbury \space
				startTexFig \space }
		\if@angle
			\special {ps:: \@p@sangle \space rotate \space} 
		\fi
		\if@clip{
			\if@verbose{
				\ps@typeout{(clip)}
			}\fi
			\special{ps:: doclip \space }
		}\fi
		\if@prologfile
		    \special{ps: plotfile \@prologfileval \space } \fi
		\if@decmpr{
			\if@verbose{
				\ps@typeout{psfig: including \@p@sfile.Z \space }
			}\fi
			\special{ps: plotfile "`zcat \@p@sfile.Z" \space }
		}\else{
			\if@verbose{
				\ps@typeout{psfig: including \@p@sfile \space }
			}\fi
			\special{ps: plotfile \@p@sfile \space }
		}\fi
		\if@postlogfile
		    \special{ps: plotfile \@postlogfileval \space } \fi
		\special{ps::[end] endTexFig \space }
		\vbox to \@p@srheight sp{
			\hbox to \@p@srwidth sp{
				\hss
			}
		\vss
		}
	}\else{
		\if@draftbox{		
			\hbox{\frame{\vbox to \@p@srheight sp{
			\vss
			\hbox to \@p@srwidth sp{ \hss \@p@sfile \hss }
			\vss
			}}}
		}\else{
			\vbox to \@p@srheight sp{
			\vss
			\hbox to \@p@srwidth sp{\hss}
			\vss
			}
		}\fi

	}\fi
}}
\psfigRestoreAt
\let\@=\LaTeXAtSign

\begin{figure}
\vspace*{-1.5 in}
\centerline{\psfig{figure=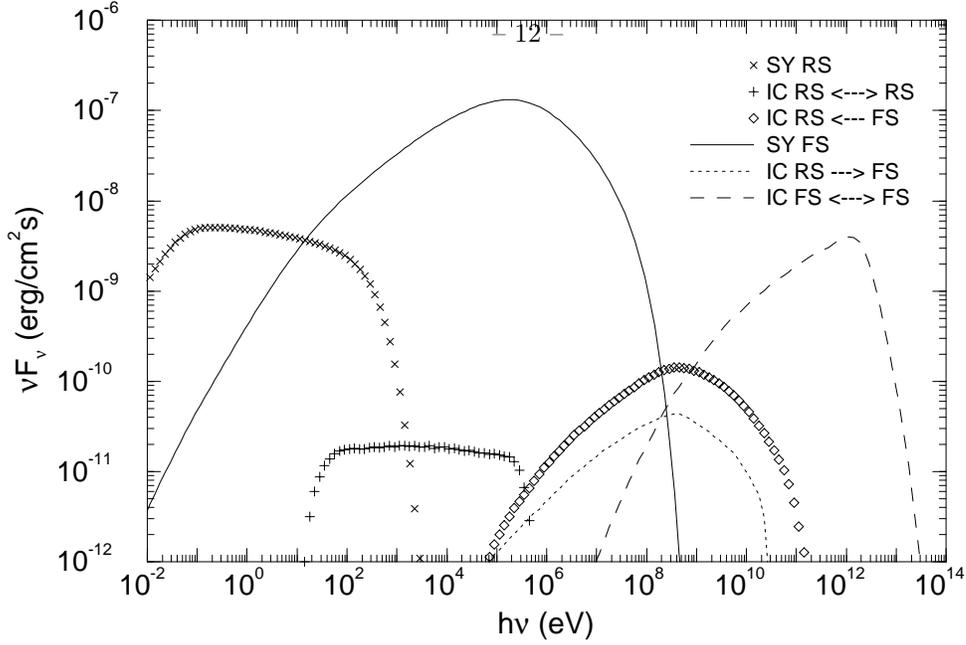}}
\vspace*{-2.4 in}
\caption{Spectrum of a GRB arising from an impulsive fireball and with 
      a weak coupling  in the radiating ejecta. The burst was obtained 
      assuming an initial energy release of  $10^{52}$ ergs/sr, and a 
      fireball with initial Lorentz factor $\Gamma_0 = 500$;
      the magnetic field, protons and electrons are at equipartition; 
      the electron distribution power-law index is $p=2.5$ and 
      $\gamma_M/\gamma_m$  is 100 for the reverse shock and 10 for the 
      forward shock. The burst is located at a redshift $z=1$. 
      The legend indicates the origin of each component,
      \eg RS $\rightarrow$ FS means reverse shock synchrotron photons 
      up-scattered behind the forward shock.}
\end{figure}

\begin{figure}
\vspace*{-0.8 in}
\centerline{\psfig{figure=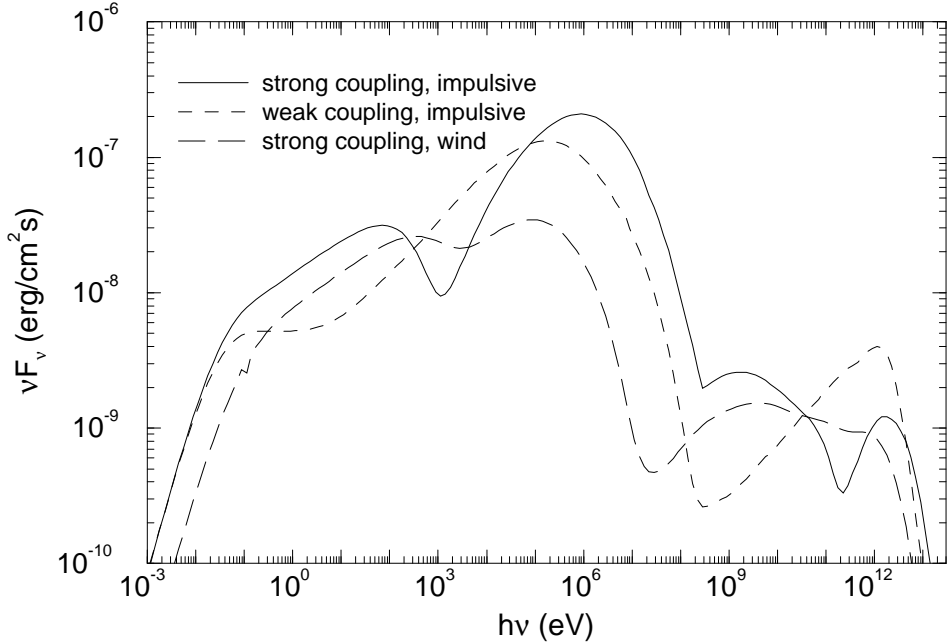}}
\vspace*{-3.1 in}
\caption{Comparison between averaged spectra obtained in three models 
      (see legend).  The burst parameters are the same as for Figure 1. 
      Note the harder and more intense burst resulting
      from an impulsive fireball and strong coupling, 
      as well as the weaker self-inverse Compton emission
      from the forward shock (around 1 TeV). 
      An extended energy release at the place where the fireball
      originates results in a softer burst, 
      in which the two shocks radiate comparable amounts of energy.}
\end{figure}

\begin{figure}
\centerline{\psfig{figure=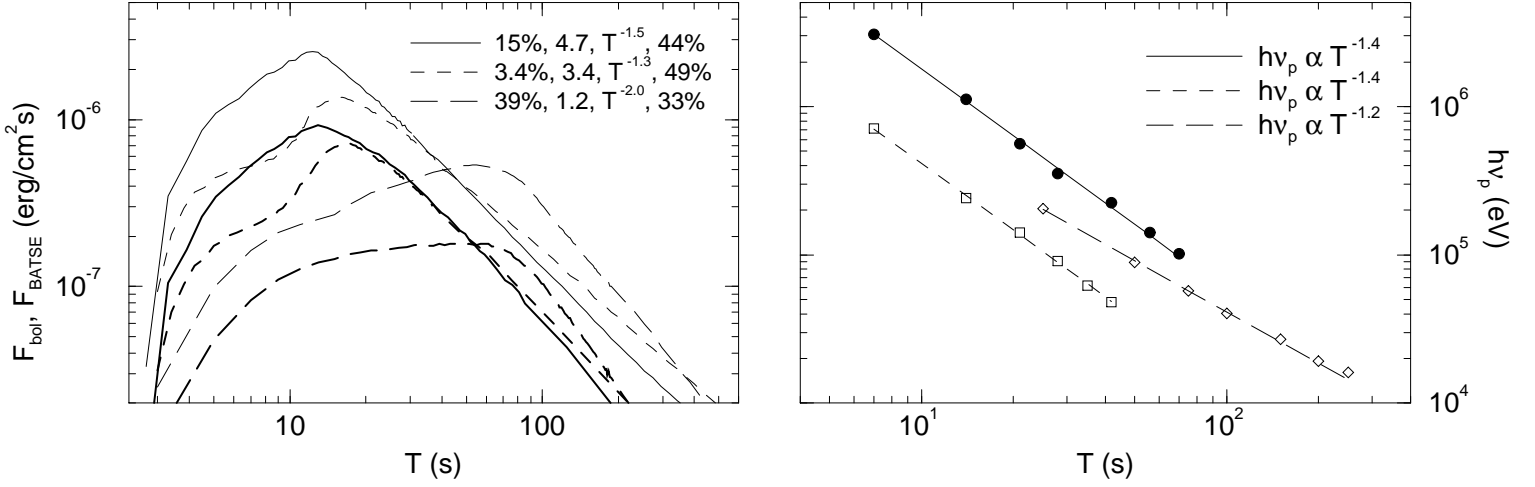}}
\vspace*{2mm}
\caption{Temporal and spectral evolution of the bursts whose spectra are 
     shown in Figure 2.  The left graph shows in log-log scale the observed 
     bolometric flux (thin curves) and the flux in the 25 keV -- 1 MeV range 
     (thick curves).
     The legend gives for each model the fractional fluence of the reverse shock,
     the light-curve temporal asymmetry, the burst $T^{-\alpha}$ fall and 
     its efficiency (see text for definitions). 
     The right graph shows the evolution of the peak of $\nu F_{\nu}$. 
     The types of lines used are the same as in Figure 2: solid for impulsive 
     and strong coupling, short dashes for impulsive and weak coupling, and long 
     dashes for wind and strong coupling.}
\end{figure}

\begin{figure}
\centerline{\psfig{figure=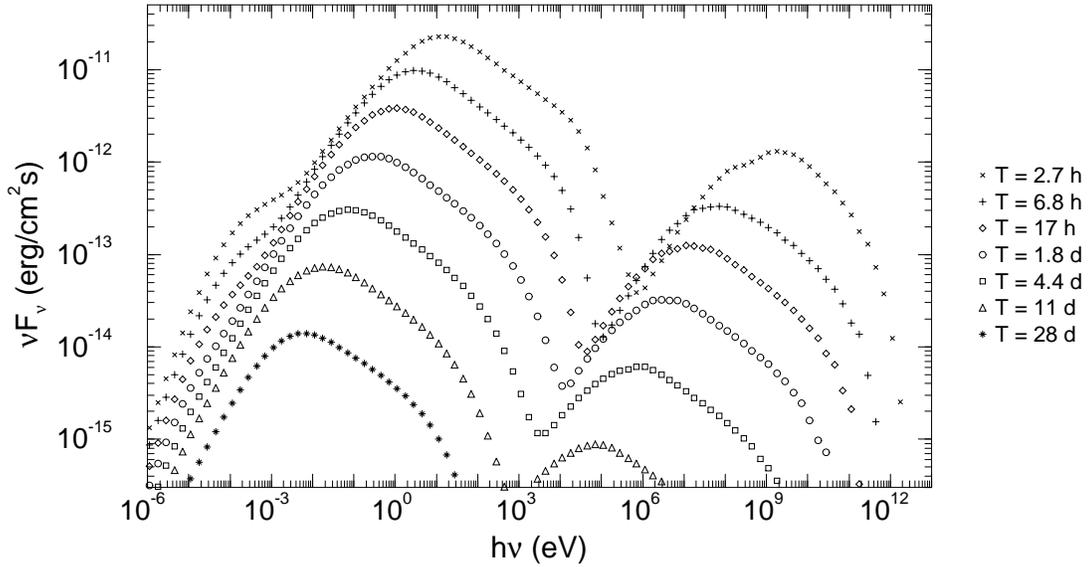}}
\vspace*{5mm}
\caption{Left graph: spectral evolution of the afterglow in the weak coupling 
     model (this is the afterglow of the GRB shown in Figure 1, but using 
     $\gamma_M/\gamma_m=100$).
     The legend indicates the observer time for each spectrum.}
\end{figure}

\begin{figure}
\centerline{\psfig{figure=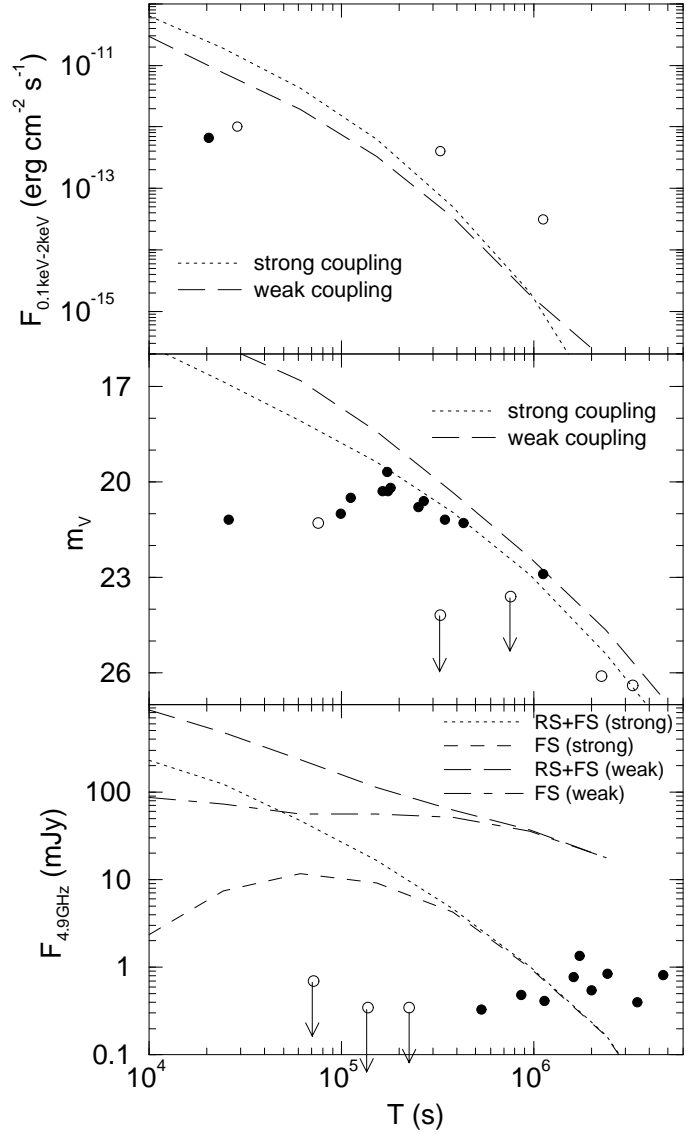}}
\vspace*{2in}
\caption{Fluence of numerically simulated afterglows in the 
    $0.1\,{\rm keV}-2\,{\rm keV}$ band, their V magnitudes and flux densities
    in radio (4.9 GHz), for both strong and weak coupling models. 
    Symbols denote real bursts: GRB 970228 (open circles)
    and GRB 970508 (filled circles), arrow showing upper limits. 
    When relevant, the fluxes from the forward shock and from both
    shocks have been shown separately (see legends of lower graphs).}
\end{figure}

\end{document}